# Remote Work: Driver or Deterrent of Digital Product Innovation

**Fangchen Song**
University of Texas at Austin
fangchen.song@mccombs.utexas.edu

**Yixuan Liu**
CEIBS
yixuanliu@ceibs.edu

**Ashish Agarwal**
University of Texas at Austin
ashish.agarwal@mccombs.utexas.edu

**Abstract**

As firms adopt divergent policies regarding work-from-home (WFH), the implications of remote work for collaborative and interdependent outcomes such as digital product innovation remain uncertain. This study examines how remote work adoption affects continuous digital product innovation using a panel dataset of mobile applications. We identify firm-level remote work adoption from job postings data and estimate its effects on app innovation using a staggered difference-in-differences design. We find that remote work significantly increases both major releases and new feature introductions per app, indicating enhanced digital product innovation performance. To assess whether these gains come at the expense of originality, we distinguish between novel and imitative feature introductions and show that remote work does not reduce the originality of digital product innovation. Moreover, improvements in digital product innovation translate into greater market success, as reflected in increased app downloads. The positive effects of remote work are stronger for app development teams with prior modular collaboration experience through open-source participation, suggesting that teams with greater experience coordinating modular work can better leverage remote work arrangements. We also find that remote work enables teams to expand their workforce and increase their collective skill capacity, both of which are associated with improved digital product innovation outcomes. In contrast, reductions in commuting time and app maturity do not explain the observed digital product innovation gains. Overall, our findings suggest that remote work can enhance continuous digital product innovation at the team level without compromising innovation novelty.

## 1. Introduction

Many organizational activities rely on teams coordinating interdependent work. The rise of remote work, enabled by digital collaboration technologies and accelerated during the pandemic, has fundamentally changed how such teams organize and execute their tasks. These changes are particularly consequential for firms developing digital products. Because digital products are created, modified, tested, and distributed largely through digital tools and infrastructures, many core development activities can, in principle, be performed remotely more readily than work tied to physical production sites or material artifacts. At the same time, digital product development depends on ongoing coordination among contributors engaged in highly interdependent tasks, making it an important setting for examining both the possibilities and limits of work-from-home arrangements. While many firms initially adopted work-from-home (WFH) policies, subsequent organizational responses have diverged. Some firms, such as Amazon and Dell, have mandated office returns, reflecting concerns that WFH may hinder collaboration-intensive activities such as digital product innovation[1]. In contrast, many other firms developing digital products continue to embrace remote or hybrid work arrangements, suggesting disagreement about its consequences. This divergence raises a central question: how does remote work affect the ability of digital product development teams to perform collaboration-intensive and interdependent innovation activities?

Prior research has primarily examined the effects of remote work on individual outcomes, such as productivity and work-life-balance (Bloom et al. 2015, Choudhury et al. 2021, Angelici and Profeta 2024, Bloom et al. 2024), as well as labor market outcomes including hiring and talent acquisition (Ham et al. 2022, Choudhury 2025, Hsu and Tambe 2025). Comparatively less is known about how remote work influences the performance of teams engaged in complex collaborative activities and the market outcomes generated by those activities. To address this gap, we examine the impact of remote work on team-level continuous digital product innovation.

This setting is particularly well-suited for studying the team-level consequences of remote work because continuous digital product innovation is inherently collaboration-intensive and interdependent, requiring sustained coordination across individuals. Continuous digital product innovation refers to the ongoing introduction and commercialization of product improvements through major releases, feature additions, and functional enhancements (Joshi et al. 2010, Gómez et al. 2017, Kahn 2018). Unlike many traditional products, digital products evolve through frequent and iterative updates, making innovation a continuous process rather than an episodic event. The ability of development teams to repeatedly introduce and commercialize product improvements is central to long-term competitive positioning (Nevens 1990, Nerkar and Shane 2007). As such, continuous digital product innovation provides a direct link between internal organizational activities and external market performance, making it a particularly informative context for assessing the team-level consequences of remote work.

[1] See companies' "return to office" announcements: https://www.aboutamazon.com/news/company-news/ceo-andy-jassy-latest-update-on-amazon-return-to-office-manager-team-ratio.
https://www.forbes.com/sites/chriswestfall/2025/02/04/rto-mandate-dell-now-requires-workers-in-the-office-5-days-a-week.

The rise of remote work has fundamentally altered how development teams organize their work, yet its consequences for digital product innovation remain poorly understood. Existing research suggests competing implications. On one hand, remote work may improve individual productivity by increasing autonomy and flexibility, enabling uninterrupted work, and reducing commuting time (Bloom et al. 2015, Choudhury et al. 2021, Angelici and Profeta 2024, Bloom et al. 2024). On the other hand, remote work may hinder the coordination required for interdependent innovation activities by limiting informal interaction, reducing rich information exchange, and making it harder to align distributed work (Brucks and Levav 2022, Yang et al. 2022, Lin et al. 2023). The net effect of remote work on digital product innovation is therefore theoretically ambiguous.

This ambiguity is especially important because digital products are often built on modular architectures (Baldwin and Clark 2000, Yoo et al. 2010). Modularity may help distributed teams divide work across relatively independent components, making remote collaboration more feasible than in settings where work is tightly interdependent (Baldwin and Clark 2000, MacCormack et al. 2006). However, modularity does not eliminate the need for integration, nor does it guarantee that teams can realize the benefits of remote work. Whether remote work ultimately enhances or hinders digital product innovation is therefore an open empirical question.

We examine this question in the context of the mobile app industry, an important setting for studying continuous digital product innovation. Innovation in this market occurs through frequent releases and feature updates that directly influence user demand and product performance (Kaushik and Gokpinar 2023, Karanam et al. 2025). The mobile app ecosystem generated nearly $1.3 trillion in billings and sales in 2024[2] and represents one of the largest digital product markets. Moreover, app development is organized around teams that continuously introduce new features and improvements. This setting therefore provides a useful context for assessing how remote work affects both innovation performance. Accordingly, we ask: how does the adoption of remote work affect innovation performance for apps? Answering this question

[2] Apple's report on the global App Store ecosystem: https://www.apple.com/newsroom/2025/06/global-app-store-helps-developers-reach-new-heights/

sheds light on how workplace flexibility shapes sustained innovation and market performance in digital product markets.

We define app innovation performance as observable outputs of continuous digital product innovation, including the number of major releases and new features introduced (Huang 2008, Foerderer et al. 2018, Kaushik and Gokpinar 2023). These measures capture the extent to which an app evolves through substantial updates and new functionalities over time. We additionally examine original new feature introductions —features that appear for the first time within a given app category— to capture innovation reflecting idea creation rather than imitation (Wang et al. 2018). To construct these measures, we collect data on version release notes, app genre classifications and rankings for iOS apps. To identify whether and when app development teams adopt remote work, we leverage job posting data from Revelio Labs and apply established classification approaches from the economics literature (Dingel and Neiman 2020, Bloom et al. 2021). Our unit of analysis is the app-month, spanning May 2020 to June 2023. Because adoption occurs at different points in time, we employ a staggered difference-in-differences design using the estimator proposed by Callaway and Sant' Anna (2021) to address biases arising from heterogeneous treatment effects. We assess robustness of our results using alternative specifications, matching approaches, and an instrumental variable (IV) strategy.

Our empirical analysis shows that remote work adoption significantly increased both major releases and new feature introductions per app, indicating improved digital product innovation performance. The results suggest that the net effect of remote work on app innovation performance is positive, with productivity gains exceeding the coordination challenges introduced by remote collaboration. While the absolute numbers of both original and imitative new features increased, the share of imitative new features remained unchanged, suggesting that the net benefits of remote work were not achieved at the expense of innovation originality. We further document that remote work adoption increased app downloads, indicating that gains in digital product innovation translated into improved market performance.

We further examine how coordination capabilities shape the relationship between remote work and app innovation performance. We find that the positive effects of remote work are significantly stronger for development teams with prior modular collaboration experience captured using open-source participation, suggesting that teams with stronger capabilities for coordinating interdependent work are better positioned to capitalize on remote work arrangements. We also find that remote work adoption is associated with growth in team size and collective skill capacity. Moreover, increases in workforce scale and collective skill capacity are positively associated with subsequent app innovation performance, suggesting that expanded access to talent is an important benefit of remote work for digital product development teams. We also find that the effects of remote work do not vary significantly across development teams with different levels of pre-treatment traffic congestion exposure or across apps with different ages, suggesting that neither reductions in commuting burden nor app maturity are primary drivers of the observed innovation gains.

Our study contributes to the literature in several ways. First, we contribute to the growing literature on remote work, which has primarily focused on individual-level outcomes such as productivity and employee well-being (Bloom et al. 2015, Choudhury et al. 2021, Angelici and Profeta 2024, Bloom et al. 2024), as well as labor market outcomes including hiring and talent acquisition (Ham et al. 2022, Choudhury 2025, Hsu and Tambe 2025). We extend this literature by examining how remote work affects the performance of digital product development teams engaged in collaboration-intensive innovation activities. By focusing on continuous digital product innovation, we move beyond individual outcomes to study an important organizational outcome that emerges from the coordinated efforts of specialized contributors. In addition, our analysis provides insight into the organizational processes through which remote work influences innovation by highlighting the roles of coordination capabilities and access to talent, while also assessing alternative explanations related to commuting burden and product lifecycle stage.

Second, our study contributes to the literature on organizational coordination and distributed innovation (Baldwin and Clark 2000, MacCormack et al. 2006). Prior research has emphasized the importance of coordination capabilities and modular organizational structures in facilitating distributed collaboration. We extend this literature by demonstrating that prior modular collaboration experience enables development teams to realize greater innovation gains from remote work. Rather than viewing remote work as uniformly beneficial or detrimental, our findings suggest that its effectiveness depends on organizations' existing coordination capabilities and their ability to manage distributed work.

Third, we contribute to the literature on digital product innovation and app development. Prior research has largely examined how product characteristics, feature design, and competitive dynamics shape user demand and market performance (Allon et al. 2022, Gu et al. 2022, Karanam et al. 2023, Deng et al. 2024, Karanam et al. 2025). In contrast, we shift attention from product and market characteristics to the organization of digital product development. Specifically, we show how remote work influences the innovation activities of development teams and, in turn, the market performance of apps.

Our findings yield several managerial implications. First, for organizations engaged in digital product development, remote work can serve as an effective organizational arrangement for supporting continuous product innovation. The positive effects of remote work on product releases, feature introductions, and downstream market performance suggest that remote work need not undermine innovation outcomes, even in settings characterized by substantial task interdependence and collaboration requirements. Managers should therefore evaluate remote work not only as a human resource policy, but also as a strategic choice that can influence the innovation performance of development teams.

Second, our findings highlight the importance of team capabilities in shaping the effectiveness of remote work. Development teams with prior experience in modular collaboration benefit more from remote work, suggesting that capabilities for coordinating modular and distributed work become increasingly valuable in remote environments. Managers seeking to capture the benefits of remote work may therefore need to invest in collaboration practices and workflows that facilitate coordination among geographically dispersed contributors.

Third, our findings suggest that access to talent is an important advantage of remote work in digital product development. By relaxing geographic constraints, remote work enables development teams to expand their workforce and increase their collective skill capacity, both of which are associated with stronger product innovation outcomes. For organizations competing in rapidly evolving digital product markets, remote work may provide access to specialized expertise and help strengthen the development capabilities needed to sustain continuous digital product innovation.

## 2. Literature Review

### 2.1 Remote Work

As remote work has become widespread, a growing body of research has examined its implications for workers and organizations. Existing studies show that remote work provides individual workers with greater autonomy and flexibility in structuring work schedules, reduces interruptions from in-person interactions, and eliminates commuting time (Bloom et al. 2014). These changes can enhance the productivity of knowledge workers (Bloom et al. 2015, Choudhury et al. 2021, Angelici and Profeta 2024). In addition, autonomy and flexibility function as nonpecuniary benefits (Choudhury et al. 2021), contributing to higher job satisfaction and improved work-life-balance (Bloom et al. 2024, Choudhury et al. 2024). However, this literature focuses primarily on individual-level outcomes and largely abstracts from interdependent work, where performance depends on coordination and communication across workers. It also does not consider how remote work policies may influence the capabilities and capacity of development teams engaged in innovation activities.

Another stream highlights potential challenges of remote work for creative and conceptual tasks. Virtual communication has been shown to reduce idea novelty (Brucks and Levav 2022), and remote scientists appear less likely to engage in ideation and manuscript writing (Lin et al. 2023). However, these studies focus primarily on ideation outputs, such as creative ideas, patents, and research publications. Such outcomes differ from continuous digital product innovation, which concerns the design, development, and release of improved products to users with the goal of creating customer value and increasing product demand (Kahn 2018). Prior research suggests that ideation is only weakly correlated with team-level product innovation (Devinney 1993, Barge-Gil et al. 2011, Gómez et al. 2017), and patents often serve strategic purposes unrelated to product delivery, such as blocking competitors or facilitating negotiations (Cohen et al. 2000, Joshi et al. 2010). Moreover, the settings examined by prior studies typically involve discrete outputs produced by relatively independent contributors, rather than the ongoing, coordinated processes through which commercial digital products are continuously improved and released.

A third stream examines hiring outcomes under remote work. By expanding the geographic scope of recruitment, remote work enables development teams to attract more experienced and diverse applicants (Hsu and Tambe 2025), and to hire suitable candidates regardless of location (Choudhury 2025). While this literature highlights how remote work changes the supply of labor, it does not evaluate how these changes translate into the performance of team-based activities such as digital product innovation.

Taken together, these literatures leave an important gap. We lack evidence on how remote work affects the performance of digital product development teams, where outcomes depend on both individual productivity and effective coordination among interdependent contributors.

We address this gap by studying the impact of remote work on continuous digital product innovation—an ongoing process through which development teams iteratively refine and upgrade products' functional characteristics, technical capabilities, and ease of use (Joshi et al. 2010, Gómez et al. 2017, Kahn 2018). Unlike discrete creative tasks, continuous digital product innovation requires repeated coordination among interdependent workers over time to develop, refine, and commercialize incremental improvements. Because the successful commercialization of innovation is central to sustained competitive advantage (Nevens 1990, Nerkar and Shane 2007), understanding how remote work affects this process is critical.

## 2.2 Mobile App Development

A growing body of research investigates the development and performance of mobile apps. Early studies highlight how app characteristics, such as pricing strategies, in-app purchases, in-app advertising, listing categories and updating strategies, shape user demand and market outcomes (Ghose and Han 2014, Lee and Raghu 2014, Comino et al. 2016). Subsequent research examines how specific features in app design, such as crowdsourcing features, social features, developer-initiated novel features and user-suggested imitative features, affect product functionalities and user engagement and retention (Gu et al. 2022, Karanam et al. 2023, 2025). A parallel stream of research explores market competition and copycat behavior in app design (Wang et al. 2018).

Despite these advances, prior research has largely focused on user-facing product attributes and feature design as determinants of market performance. In contrast, relatively little attention has been paid to the innovation process underlying digital product development or to how development teams organize and coordinate innovation activities. A small but growing literature begins to address innovation dynamics by examining how the sequencing and distinctiveness of updates relative to competitors affect app performance (Kaushik and Gokpinar 2023), and how developers balance external market feedback (e.g., user ratings) with internal constraints such as project timelines and operational limitations (Allon et al. 2022). However, even this emerging work largely abstracts from how digital product development teams are organized. In particular, although remote work has become increasingly prevalent in software development, its implications for continuous product innovation in app markets remain poorly understood.

We address this gap by examining how the adoption of remote work affects app-level continuous innovation performance and subsequent user demand in the mobile app market. By shifting attention from user-facing product attributes to the organization of development activities, our study provides new evidence on how workplace arrangements shape continuous product innovation outcomes in digital markets.

## 3. Theoretical Background

### 3.1 Remote Work and Digital Product Innovation

Prior research suggests that remote work provides employees with greater autonomy and flexibility in organizing their work, facilitates uninterrupted work periods, and frees up time that would otherwise be devoted to commuting (Bloom et al. 2014, Choudhury et al. 2021). These dynamics are particularly salient for software developers, for whom remote work has been shown to improve perceived productivity through increased focus time, schedule flexibility, and reduced interruption (Ralph et al. 2020, Ford et al. 2021) .

Furthermore, job autonomy satisfies basic psychological needs for self-direction and competence, which in turn supports intrinsic motivation and innovative work behavior, including opportunity exploration, idea generation, and idea championing (Deci and Ryan 1985, Gagné and Deci 2005, Dehkordi et al. 2025). By simultaneously improving focused execution and strengthening the autonomy conditions under which innovative behavior is more likely to occur, remote work may improve individual developers' contribution to continuous digital product innovation.

At the same time, remote work may impede coordination by limiting informal interaction, reducing the exchange of rich information, and making it more challenging to align activities across interdependent tasks (Brucks and Levav 2022, Yang et al. 2022, Lin et al. 2023). In addition, recent research finds that fully remote IT professionals experienced a decline in output per hour of 8 to 19 percent, driven primarily by rising coordination overhead and shrinking uninterrupted work time (Gibbs et al. 2023). Because digital product innovation outcomes emerge from the combined efforts of multiple contributors, increases in coordination challenges may offset productivity gains at the individual level. As a result, the overall net effect of remote work on team digital product innovation performance depends on the extent to which the benefits of increased individual productivity outweigh the costs of coordinating distributed work.

The nature of digital product development, however, suggests that this tradeoff may be more favorable than in many traditional innovation settings. Digital products are typically built using modular architectures in which complex systems are decomposed into relatively independent components connected through standardized interfaces and design rules (Baldwin and Clark 2000). By allowing work to be partitioned into smaller and more self-contained tasks, modular architectures of digital products reduce coordination requirements and permit parallel experimentation among contributors (Baldwin and Clark 2000). These characteristics are particularly salient in software and app development, where development teams frequently work on distinct features, functionalities, or code modules that can be designed and tested independently before being integrated into the broader product (Parnas 1972, MacCormack et al. 2006). Consequently, although remote work may increase coordination challenges, the modular nature of digital product development reduces the extent to which such innovation depends on constant face-to-face interaction. In line with this argument, recent evidence suggests that the impact of virtuality on innovation depends primarily on task design, with low interdependence and well-structured work representing the most important moderating conditions (Sinnemann and Weiss 2025).

Taken together, these arguments suggest that for digital product development teams working on modular, component-based innovation, the structural properties of the task environment lower the threshold at which remote work's productivity and autonomy benefits can outweigh its coordination costs. Whether teams can fully realize those benefits in practice, however, depends on the degree to which they actually operate in the low-coordination-cost manner that modular architecture makes possible.

### 3.2 Remote Work and Modular Collaboration Experience

Although modular architectures reduce coordination demands, teams differ in their ability to capitalize on this structure. Modularity makes it possible to decompose work, standardize interfaces, and support parallel development, but these advantages are not automatic. They depend on whether team members have learned how to coordinate within such an architecture. Teams with greater modular collaboration experience are more likely to understand component boundaries, interface conventions, and integration routines, allowing them to work with fewer dependencies and less ongoing coordination. Thus, modular collaboration experience reflects a team's capacity to convert the structural affordances of modularity into realized coordination efficiencies.

The autonomy and productivity benefits of remote work are more likely to improve team-level innovation when coordination overhead remains manageable. The production of innovation requires the integration of differentiated knowledge and expertise contributed by multiple individuals toward the creation of new product functionality (Kleis et al. 2012). High-quality coordination enables teams to integrate these heterogeneous inputs, align interdependent tasks, and sustain collaborative problem solving, thereby enhancing the effectiveness of collective work (Hoegl et al. 2004, Majchrzak et al. 2005). Prior research further suggests that effective coordination depends on established routines, communication structures, and shared understandings that facilitate information exchange and knowledge integration among contributors (Majchrzak et al. 2005, Enkel et al. 2009, Kleis et al. 2012). Teams with greater prior experience coordinating modular work are more likely to have developed routines and shared understandings that reduce reliance on informal face-to-face interaction, allowing them to function more effectively under remote conditions.

These capabilities are particularly valuable in digital product development, where innovation frequently emerges from the parallel development and integration of multiple features, functionalities, and software components (Parnas 1972, MacCormack et al. 2006). Teams with deeper modular collaboration experience should be better able to preserve the coordination advantages of modularity under remote work and, in turn, translate WFH's individual-level benefits into collective digital product innovation outcomes. Accordingly, if WFH improves digital product innovation by lowering coordination costs in modular systems, its effect should be stronger for teams with greater prior modular collaboration experience.

### 3.3 Remote Work and Development Team Capabilities

Beyond its implications for individual productivity and autonomy, remote work policy may also shape continuous digital product innovation through team-level capability mechanisms. In particular, by relaxing geographic constraints on hiring, remote work can alter both the size and the knowledge composition of development teams. These represent distinct but complementary pathways through which remote work policy may enhance innovation performance, extending the individual-level mechanisms discussed above.

*Team Size*: Remote work may increase development team size through both resource reallocation and labor market access mechanisms. First, by reducing expenditures associated with physical office space and related facilities, organizations may be able to reallocate resources toward development headcount (Choudhury 2025). Second, by decoupling hiring from geographic proximity, remote work expands the set of potential contributors available to the firm, thereby increasing the feasibility of staffing larger development teams (Choudhury 2025).

Team scale is consequential for continuous digital product innovation because larger development teams can allocate more labor to product development activities, pursue parallel experimentation, and support a broader portfolio of concurrent technical initiatives (Cusumano 1997, Mao et al. 2016). Prior research similarly suggests that the number of individuals engaged in knowledge creation is an important determinant of innovation output, as larger groups bring greater aggregate effort and task capacity to innovation activities (West and Anderson 1996, Bloom et al. 2020). Larger teams are often more effective at developing, extending, and refining existing ideas (Wu et al. 2019). Thus, remote work may enhance digital product innovation by enabling development teams to grow beyond the scale constraints imposed by local labor markets.

*Team Skill Capacity*: Remote work may also enhance the collective skill capacity of development teams. When hiring is constrained by geographic proximity, organizations may face limitations in the range of expertise available within local labor markets. By relaxing these constraints, remote work enables organizations to recruit individuals possessing specialized technical knowledge and experiences that may not be locally available (Choudhury 2025). As a result, development teams may accumulate a broader set of skills and technical capabilities.

Prior research suggests that greater breadth of knowledge and expertise enhances organizational learning and innovation capability (Kogut and Zander 1992). Teams drawing on a broader range of knowledge, experiences, and specialized expertise are better positioned to recombine existing knowledge, address complex technical challenges, explore alternative solution paths, and generate novel ideas (Ancona and Caldwell 1992, Taylor and Greve 2006, Choudhury and Haas 2018, Ferrucci and Lissoni 2019). For digital product development teams, greater collective skill capacity expands the range of technical challenges that can be addressed internally and increases the team's ability to develop, implement, and refine new product features. Consequently, remote work may enhance digital product innovation by increasing the collective skill capacity available within app development teams.

## 4. Data

To examine the impact of remote work on app continuous innovation performance, we construct an app-month panel of iOS apps drawn from five major App Store categories: education, games, lifestyle, business, and utilities. The sample includes apps that appeared at least once in the daily top 500 ranking within their category between 2018 and 2023[3]. These categories represent the most popular app types, with national statistics indicating that over 70% of mobile users have apps from these categories[4]. They capture diverse and essential user needs and are thus well-suited for studying general innovation patterns. For each app, we collect version release notes, genre, and ranking information from May 2020 to June 2023 from a leading mobile app analytics provider.

[3] Although our remote work data are available only from 2020 onward, app rankings change over time. We therefore use top 500 rankings over the 2018–2023 period to identify the apps included in the sample, which provides broader coverage of relevant apps.

[4] https://www.statista.com/statistics/270291/popular-categories-in-the-app-store/.

We focus on development teams which are responsible for the digital product innovation. To identify whether and when development teams adopted remote work, we use job posting data from Revelio Labs, a source widely used to study organizational characteristics and workplace practices (Gutiérrez et al. 2020, Eisfeldt et al. 2023, Campello et al. 2024, Goldfarb et al. 2025). We focus on 11 development-related roles, including application engineer, software engineer, QA tester, project manager, product manager, data analyst, IT specialist, infrastructure engineer, technical architect, IT project manager, and technician. Prior research shows that both technical specialists and project- or product-oriented roles are central to designing functionality and overseeing execution; accordingly, these roles capture the core personnel responsible for developing, implementing, testing, and managing digital product updates (Jalil and Shahid 2008, Langer et al. 2014, Kalliamvakou et al. 2017, Venkatesh et al. 2018). Focusing on these roles allows us to measure remote work adoption specifically within the workforce engaged in digital product innovation.

To determine the timing of remote work adoption, we follow established approaches in the economics literature (Dingel and Neiman 2020, Bloom et al. 2021, Ham et al. 2022) and search for job titles and descriptions for WFH-related keywords. Because such terms may also appear in restrictive statements, for example, indicating that remote work is not permitted, we apply a rule-based text classification procedure that evaluates the local linguistic context surrounding each keyword. We define the adoption date as the first month in which a permissive statement indicates that remote work is allowed. Additional details on the classification procedure are provided in Appendix A.

To connect app dataset with labor market dataset, we follow established record-linkage approaches that reconcile organization name variations through standardization and string-matching techniques (Tambe and Hitt 2012). Specifically, we merge the app-level dataset with Revelio Labs job posting data using a contextual name-matching procedure that links app developer names to company names. This process yields an initial sample of 3,626 apps and their associated development teams. In addition, we focus on apps whose development teams consist of at least three members. This restriction ensures that the sample captures settings in which digital product development relies on collaboration among multiple contributors and therefore provides an appropriate context for examining how remote work affects collaboration-intensive innovation activities within digital product development teams. We then construct a monthly panel from May 2020 to June 2023.

Our main analysis uses a staggered difference-in-differences design that exploits variation in the timing of WFH adoption across apps. To support this identification strategy, we retain only apps whose associated development teams adopt remote work during the sample period but at different points in time. Apps associated with development teams that indicate remote work from the beginning of the sample period are excluded because they lack the pre-treatment observations necessary for identification. We also exclude apps associated with development teams that never adopt remote work during the sample period, such that identification is based on comparisons between earlier and later adopters. The final sample consists of 1,782 apps with valid pre-treatment periods, observed remote work adoption during the panel, and development teams containing at least three members. Treatment is defined as the first month in which an app development team publishes a job posting containing a permissive remote-work statement. Months prior to adoption are classified as pre-treatment, and months following adoption as post-treatment. As discussed in Section 6.2, we conduct robustness analyses using alternative thresholds for defining treatment timing and assigning apps to treatment and control groups.

## 5. Empirical Analysis

### 5.1 Measures

In the mobile app industry, it is standard practice to publish release notes alongside each version update to summarize the changes implemented (Kaushik and Gokpinar 2023). Prior research shows that these release notes reflect underlying software development activities and provide detailed information about both the scope and novelty of product changes (Baysal and Malton 2007, Yu 2009, Foerderer et al. 2018). We therefore leverage this feature to construct panel measures of app-level innovation output.

Our first primary outcome is the number of major releases in each app-month. We classify 98,512 historical version logs from the selected apps using an LLM-based approach. Following the literature, we code a release as *major* when it introduces fundamental changes, such as substantial redesigns of core functionality or the addition of significant capabilities, and as *minor* when it primarily reflects maintenance activities, including bug fixes and performance tuning that do not significantly change the app's core functionality (Foerderer et al. 2018). This classification enables us to distinguish between substantive product innovation and routine maintenance updates. To operationalize this distinction, we prompt OpenAI's GPT-5.5 model, which is well suited for interpreting the nuanced and often unstructured descriptions in software release notes (Huang et al. 2026). Online Appendix B provides additional details on the prompting procedure and human validation.

We also consider the introduction of new features as a measure of innovation outcome. Unlike the major release measure, which evaluates the overall scope of an update, the new feature measure focuses directly on the introduction of novel functionalities. We identify features from release-note text using embedding-based semantic clustering (OpenAI's text-embedding-3-large model) and classify a feature as *new* if the underlying terminology has not appeared in any prior release of the same app following established approaches (Huang 2008, Kaushik and Gokpinar 2023); features reusing previously observed terminology are classified as *existing*. Online Appendix B provides additional details on feature extraction and human validation. On average, each major release introduces approximately 6.3 new features, suggesting that major releases typically bundle multiple novel functionalities. This pattern provides validation for our classification approach, as releases involving substantive product changes are also those most likely to incorporate new capabilities. We therefore use both major releases and new feature introductions as measures of app innovation performance. It is possible that some features are not original but are copied from competitors. In order to verify our results for original features, we further classify new features as *original* if they appear for the first time within an app category and as *copycat* otherwise.

Table 1 reports variable definitions and summary statistics for the app-month variables used throughout the empirical analyses.

[insert Table 1 here]

### 5.2 Model and Estimation

Our objective is to examine the impact of remote work adoption on app innovation performance. As described above, we measure app innovation performance using the number of major releases and the number of new features. Identifying the impact of remote work adoption on app innovation performance poses several empirical challenges. First, apps may differ in persistent but unobserved characteristics that influence both the likelihood that their development teams adopt remote work and their innovation outcomes. Examples include product characteristics, product architectures, and relatively stable attributes of the development organization that shape innovation activities. Second, remote work adoption occurs at different points in time across development teams, creating a staggered treatment structure under which conventional two-way fixed effects (TWFE) estimators may yield biased estimates in the presence of heterogeneous treatment effects (Goodman-Bacon 2021).

To address these challenges, we implement a staggered difference-in-differences (DiD) estimator proposed by Callaway and Sant' Anna (2021) (hereafter CS). The CS estimator identifies the policy effect by decomposing the staggered adoption design into a series of two-period comparisons between treated and control groups. Let $G_i$ denote the period in which the developers of app $i$ first adopt remote work. For units with $G_i = g$ and calendar time $t \geq g$, the CS estimator first estimates the counterfactual common trend between the reference period $g-1$ (one period prior to adoption) and calendar time $t$ using observations from the control group:

$$m_{g,t}(x) = \mathrm{E}\left[Y_{it}(0) - Y_{i,g-1}(0) \,\middle|\, X_i = x\right] \quad (1)$$

The group-time average treatment effect on the treated (ATT) is then computed by subtracting the estimated common trend from the temporal changes in $Y$s among the treatment group where $G_i = g$:

$$ATT(g,t) = \mathrm{E}\left[Y_{it} - Y_{i,g-1} - m_{g,t}(X_i) \,\middle|\, G_i = g\right] \quad (2)$$

which allows treatment effects to vary across adoption cohorts $g$ and over calendar time $t$. The dependent variable $Y_{it}$ represents the log number of major releases and the log number of new features of app $i$ in month $t$. We take the log transformation to reduce the skewness of our outcome variables. In our analysis, we implement the CS estimator using the doubly robust estimator with not-yet-treated apps as the control group. The CS estimator compares changes in outcomes for treated apps with changes for apps that have not yet adopted remote work in the same calendar periods, thereby accounting for time-invariant app heterogeneity and common time shocks while allowing treatment effects to vary across adoption cohorts and event time.

In the results section, we present event-study plots that aggregate ATT estimates by event time $e = t - g$. Our baseline event window spans 12 months before and 12 months after each app's adoption of remote work. As presented in the robustness check section, our results remain consistent to alternative window lengths, such as 6 or 18 months.

***Identification***: Our identification relies on a cohort-specific parallel trends assumption, which requires that, in the absence of remote work adoption, apps adopting remote work in period $g$ would have followed the same average outcome trends as apps that had not yet adopted remote work by period $t$. Using not-yet-treated apps as the control group, the CS estimator compares changes in outcomes between treated and control apps within the same calendar periods. Through this difference-in-differences design, the estimator accounts for time-invariant app-specific heterogeneity by comparing changes in outcomes relative to each app's pre-treatment period, thereby differencing out persistent factors such as product characteristics that shape innovation activities. By comparing treated and not-yet-treated apps within the same calendar periods, the estimator also accounts for common temporal shocks, including industry-wide technological trends and platform-level changes affecting app-level innovation outcomes. The approach further accommodates staggered treatment timing and heterogeneous treatment effects across adoption cohorts and event time.

We conduct several additional analyses to address potential concerns regarding identification and measurement. First, to ensure that our treatment definition captures meaningful adoption of remote work within development teams, we refine the timing of treatment. In the baseline specification, treatment begins in the first month in which the share of job postings that allow remote work becomes positive. In the alternative specification, we instead define treatment start time as the first month in which the share of job postings that allow remote work exceeds the median in the sample. Apps that have not yet crossed this median threshold are reassigned to the control group. This approach allows us to focus on more substantive adoption of remote work within development teams.

Second, to mitigate concerns that remote work adoption is nonrandom, we perform analyses using look-ahead propensity score matching (LA-PSM) and two-way fixed effects (TWFE) models. Third, to disentangle the effects of remote work from broader pandemic-related shocks, we replicate the analysis using only post-pandemic observations.

Fourth, we address endogeneity concerns, including potential measurement error in the treatment variable, reverse causality between app innovation outcomes and remote work adoption, and omitted-variable bias arising from unobserved shocks, by implementing an instrumental variable (IV) strategy. Fifth, we examine an alternative outcome variable to ensure that our results are not driven by a specific operationalization of app innovation performance.

Sixth, we account for potential policy fluctuations by excluding observations in which app development teams change their remote work policies. Seventh, we consider the temporal dynamics of remote work adoption by examining lagged effects on app innovation performance and conducting analyses at the quarterly level. Finally, we aggregate app innovation outcomes to the firm level to assess whether our findings extend beyond individual apps to overall firm innovation performance. Details of these robustness analyses are provided in Section 6.2.

## 6. Results

### 6.1 Main Results

Table 2 reports the estimated effects of remote work adoption on app innovation performance. Consistent with our theoretical argument that remote work enhances continuous digital product innovation, we find that the adoption of remote work policies is associated with significant improvements in app innovation outcomes. Specifically, remote work adoption is associated with a 10.0% increase in major releases and an 11.1% increase in new feature introductions at the app level. To evaluate the validity of the parallel trends assumption, we plot event-time estimates for the main outcome variables in Figure 1. The absence of significant pre-trends suggests no differential trends prior to adoption, supporting the credibility of our identification strategy.

These results provide empirical support for the view that remote work, as an organizational shift, can improve continuous innovation performance of digital product development teams. Beyond user-side and market determinants documented in prior research (Allon et al. 2022, Gu et al. 2022, Karanam et al. 2023, Deng et al. 2024, Karanam et al. 2025), internal work arrangements can also meaningfully influence app innovation performance. Also, in the context of remote work, where coordination frictions and communication costs may increase, the observed improvement in app innovation outcomes indicates that the productivity gains of remote work outweigh the coordination challenges for app development teams.

[insert Table 2 here]

[insert Figure 1 here]

### 6.2 Robustness Checks

To further investigate the robustness of our findings, we employ a range of different tests to examine the relationship between remote work adoption and app innovation performance.

***Treatment Specifications:*** In our main analysis, we define the adoption date as the first month in which a job posting for roles relevant to app development explicitly allows remote work. To address the concern that remote work observed in only a small subset of development-related positions may not reflect a broader shift in work arrangements, we implement a stricter treatment definition. Specifically, we define remote work adoption as occurring when the share of app development positions permitting remote work exceeds the median share across development teams. The earliest month in which this threshold is satisfied is designated as the treatment start time. Using a 12-month event window before and after adoption and employing later adopters as the control group, we re-estimate our baseline specification. As reported in column 1 of Tables 3 and 4, remote work adoption increased both major releases and new feature introductions. These findings are consistent with our main results and suggest that our conclusions are not driven by marginal or isolated instances of remote work availability within app development teams.

***Window Size:*** To examine the sensitivity of our findings to the choice of event window, we re-estimate the baseline specification using alternative windows of 18 months before and after treatment start time, as well as 6 months before and after treatment start time. The results using the 18-month window are reported in column 2 of Tables 3 and 4, and those using the 6-month window are presented in column 3 of Tables 3 and 4. Across both alternative specifications, remote work adoption increased major releases and new feature introductions. These findings are consistent with the results obtained under the 12-month window and indicate that our conclusions are not sensitive to the choice of event window length.

***LA-PSM and TWFE:*** To address concerns regarding the comparability of early-treated and later-treated units and the potential contamination of control groups by units that will be treated in the near future, we implement a look-ahead propensity score matching (LA-PSM) approach combined with a two-way fixed effects (TWFE) model. LA-PSM matches treated apps with comparable control apps based on pre-treatment characteristics while accounting for future treatment timing (Bapna et al. 2018). We define the first month in which remote work is adopted within development teams as the treatment start time. To construct matched controls, we estimate propensity scores using each app's pre-treatment download volume, pre-treatment major release count, and pre-treatment new feature count, and perform one-to-one nearest-neighbor matching without replacement using a caliper of 0.2 standard deviations of the logit propensity score (Austin 2011). This yields 819 treated apps matched to 819 control apps, resulting in a sample of 1,638 apps. Column 4 of Tables 3 and 4 report the corresponding estimates. The results show that remote work adoption increased both major releases and new feature introductions, consistent with our main findings. We report parallel trends tests for the matched sample in Table C.1 of Online Appendix C, which support the validity of the identification strategy.

***Post-pandemic Analysis:*** During the pandemic, remote work was mandated for many organizations. Prior research shows that the pandemic had a negative effect on innovation outcomes. For example, firms reduced innovation spending not only in the first year of the pandemic but also in the two subsequent years (Trunschke et al. 2024), and scientists initiated significantly fewer new research projects during this period (Gao et al. 2021). These findings suggest an overall decline in innovative activity during the pandemic, and it is reasonable to assume that continuous digital product innovation may have been similarly depressed. At the same time, the pandemic may have introduced countervailing forces that could increase digital product innovation, such as heightened user demand for digital products and services, which could have incentivized development teams to accelerate digital product updates.

To address the concern that the observed improvement in app innovation performance may be driven by pandemic specific factors rather than remote work, we exclude pandemic period observations using the pandemic stringency index (Hale et al. 2021, Roser 2021), which measures the strictness of government policies such as lockdowns, where higher values indicate stricter policies. For each app development team, we identify the end of the pandemic period as the time when the stringency index remains below 50, and we drop all observations prior to this point. As reported in column 5 of Tables 3 and 4, remote work adoption increased both major releases and new feature introductions. These findings indicate that the observed improvements in app innovation performance are unlikely to be primarily driven by pandemic-related shocks.

[insert Table 3 here]

[insert Table 4 here]

***Instrumental Variable:*** To further address endogeneity concerns, we implement an instrumental variable (IV) analysis as a robustness check. These concerns include omitted variables, such as unobserved shocks that may affect app innovation performance; potential measurement error in identifying remote work adoption from job postings; and reverse causality, whereby app innovation outcomes may influence a development team's decision to adopt remote work. The IV approach helps mitigate these concerns by isolating variation in focal development teams' remote work adoption that is predicted by pre-treatment digital infrastructure conditions among indirectly connected labor-market peers.

Following prior work that leverages exogenous attributes of non-overlapping peer groups (Fan 2013, Karanam et al. 2025), we construct an instrument based on the lagged internet connectivity in the locations of focal development team's labor competitors' competitors. The key idea is that internet connectivity affects the feasibility and attractiveness of remote work for indirectly connected teams, and that these remote-work decisions propagate through the labor-competition network even if there is no product competition.

Consistent with prior work that conceptualizes the economy as comprising distinct labor markets segmented by occupation, industry, and geography (Şahin et al. 2014), we define a labor market as the set of workers within a given segment and the app development teams recruiting from that segment. Using Revelio Labs data, we construct labor markets based on observed labor flows among development teams. This allows us to identify teams that compete for similar talent, even when they do not directly compete in product markets. Figure 2 illustrates the logic. Consider a focal development team $A$ that competes for labor with team $B_1$ in labor market $M_1$. Team $B_1$, in turn, competes with teams $C_1$ and $C_2$ in a different labor market $M_2$. We define $B_1$ as a first-layer labor competitor of $A$, and $C_1$ and $C_2$ as second-layer competitors (i.e., competitors of $A$'s competitors). By construction, the $C$-teams are not directly connected to $A$ through labor flows.

[insert Figure 2 here]

To strengthen the exclusion restriction, we impose two conditions. First, the focal team $A$ and both the first-layer competitors $B_i$ and the second-layer competitors $C_i$ must operate in different industries and geographic locations, ensuring no direct competition in product markets. Second, there must be no direct labor flows between $A$ and $C_i$, ruling out direct labor market competition. These restrictions ensure that the $C_i$ teams affect $A$ only indirectly through shared connections with first-layer competitors $B_i$. We then measure internet connectivity in the locations of the $C_i$ teams using Ookla NetIndex data (Sood 2017) and aggregate this measure through the labor competition—first from $C_i$ to $B_i$, and then from $B_i$ to $A$—using pre-treatment labor flow shares as weight. To account for location-specific factors that may influence $A$'s remote work decision, such as regional infrastructure or local technology ecosystem characteristics, we include location fixed effects for $A$ as controls. To further address the possibility that changes in $B_i$'s labor resources affect $A$ through channels other than remote work adoption, we additionally control for the lagged number of employees and lagged skill composition of $B_i$.

The instrument is relevant for two reasons. First, internet connectivity reflects local digital infrastructure, which directly affects the feasibility of remote work. Second, because remote work provides flexibility benefits to employees (Bartik et al. 2025), development teams are more likely to adopt remote work policies when facing greater labor competition (Ham et al. 2022). Improved connectivity in the locations of $C_i$ increases their adoption of remote work, intensifying labor competition faced by $B_i$. In response, $B_i$ adjust their own remote work policies to attract and retain talent. Because $A$ competes with $B_i$ in the labor market $M_1$, these adjustments propagate to $A$'s remote work adoption decision. Thus, variation in internet connectivity among second-layer competitors generates exogenous variation in the focal team's remote work decision.

The exclusion restriction is also plausible. By construction, $A$, $B_i$, and $C_i$ are in different locations, industries, and product markets, so product strategy, release timing, and product-market innovation responses are unlikely to directly affect $A$'s innovation outcomes. In addition, $A$ and $C_i$ are not directly connected in the labor market, so $C_i$'s connectivity should not directly affect $A$'s labor market. Moreover, internet connectivity is determined by regional infrastructure and is unlikely to be influenced by team-level innovation or employment decisions. As we account for $B_i$'s labor resources such as number of employees and skill composition, it is less likely that $C_i$'s connectivity directly affects A's labor market via $B_i$'s labor resources.

We report the instrumental variable results in Table 5[5]. In the first stage, internet connectivity in the locations of the $C$ teams is strongly and positively associated with remote work adoption by focal team $A$. The corresponding F-statistic exceeds 74, indicating that the instrument is highly relevant. This pattern suggests that remote work adoption by the focal team responds to labor market pressure propagated through competitors' competitors. In the second stage, remote work adoption significantly increased both major releases and the introduction of new features, consistent with the main results.

[5] The IV estimation sample differs from the main analysis sample because the instrument is only defined for focal development teams with labor competitors' competitors that satisfy our identification criteria. Specifically, the focal team ($A$), its first-layer labor competitors ($B_i$), and the second-layer labor competitors ($C_i$) must operate in different industries and geographic locations, and there must be no direct labor flows between $A$ and $C_i$. Applying these restrictions results in an estimation sample of 633 apps.

[insert Table 5 here]

***Alternative measure:*** To address potential bias arising from the use of large language models to identify major releases and new features from app release notes, we supplement our text-based measures with an alternative approach grounded in semantic versioning conventions. Following prior research (Kemerer and Slaughter 2002, Yu 2009, Boudreau 2012, Raemaekers et al. 2014, Tiwana 2015), we distinguish major and minor updates based on the numerical versioning scheme commonly adopted in software development (e.g., 2.0.0, 2.1.1). Specifically, we classify version numbers using the "Major.Minor.Patch" structure, where an increment in the major digit reflects substantial changes or the introduction of new functionality. We count the number of major version increments based on this convention and replicate our main analysis using this alternative measure of app innovation output. The results, reported in Table C.2 of Online Appendix C, are consistent with our primary findings.

***WFH Policy Fluctuation:*** To address the concern that remote work policies may change over time, we re-estimate our results after excluding periods in which remote work adoption appears to decline substantially. For each development team, we compute the monthly share of job postings that allow remote work. We then drop observations after months when this share decreases by more than 50 percentage points relative to the prior level, which may indicate a meaningful reduction or discontinuation of remote work flexibility. We report the results in Table C.3 of Online Appendix C and find patterns consistent with our main analysis.

***Lagged Effects:*** To address the concern that the observed effects of remote work primarily reflect contemporaneous impacts, we examine whether these effects materialize with a delay. Specifically, we estimate a lagged treatment specification in which remote work adoption in period $t$ is used to explain major releases and new feature introductions in period $t + 1$. As shown in columns 1 and 2 of Table C.4 in Online Appendix C, the estimated effects remain positive and statistically significant and are quantitatively similar to our baseline results. These findings suggest that the impact of remote work on continuous app innovation is not driven solely by contemporaneous timing assumptions.

We further address the possibility that short-term monthly fluctuations introduce noise into the estimation by aggregating the data to the quarterly level and re-estimating the lagged treatment model. As reported in columns 3 and 4 of Table C.4 in Online Appendix C, the results remain consistent with our main findings, indicating that our conclusions are robust to alternative temporal aggregation.

***Firm-level Performance:*** To address the concern that innovation outcomes measured at the individual app level may not fully capture firm-level innovation performance, we aggregate app innovation outcomes to the firm level and re-estimate the impact of remote work on overall firm innovation. The results are reported in Table C.5 of Online Appendix C. We find that remote work adoption significantly increased the total number of major releases and new features at the firm level. These findings suggest that the positive effects of remote work extend beyond individual apps and translate into improvements in overall firm product innovation performance.

### 6.3 Innovation Originality and App Demand

Although we document increases in major releases and new features following remote work adoption, the baseline new feature measure is defined relative to each app's own historical releases. It is therefore important to assess whether the additional features reflect original innovation or simply imitation of existing features already introduced elsewhere in the market. Given the prevalence of copycat behavior in the app industry (Wang et al. 2018), we distinguish between original and imitative innovation using a cross-app comparison approach. Specifically, we classify the first app within a category to introduce a given feature as the original innovator. A new feature is considered *original* if it appears for the first time in the focal app and has not previously appeared in any other app in the same category. In contrast, a new feature is classified as *copycat* if it is new to the focal app but has already been introduced by another app in the same category. Based on this classification, we construct three outcome variables: the number of original new features, the number of copycat new features, and the share of copycat new features among all new features. As reported in columns 1 to 3 of Table 6, remote work adoption increased both original and copycat new features.

However, the proportion of copycat new features relative to total new features remained unchanged. These findings suggest that remote work increased overall app innovation output without reducing the originality of app innovation.

We next examine whether the improved app innovation performance affects app demand, which is particularly important given the competitive nature of the app market. To assess app demand, we use App Store ranking data. Following prior literature (Brynjolfsson et al. 2003, Chevalier and Goolsbee 2003), we assume the relationship between app sales rank and download volume follows a Pareto distribution. Using established methods validated in previous studies (Ghose et al. 2006, Garg and Telang 2013, Ghose and Han 2014), we estimate the shape and scale parameters and use them to compute demand for each app. As shown in column 4 in Table 6, remote work adoption led to a significant increase in app download volume. This evidence indicates that improvements in continuous digital product innovation under remote work are accompanied by stronger market demand.

[insert Table 6 here]

### 6.4 Remote Work and Development Capabilities

#### 6.4.1 Modular Collaboration Experience

To examine whether prior modular collaboration experience shapes the effects of remote work, we identify organizations that participated in Open-Source Software (OSS) projects on GitHub prior to treatment. We focus on OSS projects because they are characterized by modular architectures and development processes that reduce coordination costs and support distributed collaboration (MacCormack et al. 2006, Howison and Crowston 2014, Lindberg et al. 2016). We use such OSS development experience as a proxy for prior modular collaboration experience and construct a binary indicator based on whether an app development team possesses such experience before treatment. We then split the sample into two groups according to this indicator and re-estimate our main model separately for each subgroup. Table 7 reports the results. Although remote work increased app innovation performance for both groups, the increase was significantly larger among teams with prior modular collaboration experience. This finding is consistent with our arguments that prior experience coordinating work in modular development environments helps teams adapt more effectively to remote work arrangements by further reducing coordination frictions and facilitating the integration of distributed contributions. As a result, these teams are able to realize greater digital product innovation gains following the adoption of remote work.

[insert Table 7 here]

#### 6.4.2 Development Team Size and Skills

By broadening the geographic scope of the talent pool and reducing expenditures on physical office space (Choudhury 2025), remote work can free resources that may be reallocated toward hiring additional personnel. A larger workforce devoted to digital product innovation can increase total labor capacity, support parallel experimentation, and facilitate the execution of complex implementation tasks, thereby strengthening product innovation capability (Cusumano 1997, Mao et al. 2016). To examine this explanation, we analyze the relationship between remote work adoption and team size.

To isolate changes in team size from concurrent changes in team skills or organizational structure, we focus on headcount growth within pre-existing role–skill categories. Specifically, we first identify all distinct role–skill combinations observed before treatment, where a role–skill combination is defined as a specific organizational role paired with a particular skill profile. For each team $i$ that adopts remote work at time $T_i$, we construct the set of role–skill combinations that existed before $T_i$. We then track changes in the number of individuals associated with each of these pre-existing role–skill combinations over time. This approach captures workforce expansion while holding the composition of roles and skills constant, allowing us to isolate the scale effect of remote work from other changes in team composition. As reported in column 1 of Table 8, remote work adoption is associated with a significant increase in the number of individuals within pre-existing role–skill categories. This finding is consistent with the argument that remote work facilitates workforce expansion, which may subsequently enhance development capacity and digital product innovation output.

Beyond expanding team size, remote work may also affect the collective skill capacity of development teams. By relaxing geographic constraints on hiring, remote work enables development teams to recruit qualified employees from a broader talent pool (Choudhury 2025), potentially increasing access to skills that may be difficult to acquire under location-constrained hiring arrangements. Prior research suggests that broader knowledge and expertise enhance organizational learning and innovation capability (Kogut and Zander 1992). Development teams drawing on a wider range of knowledge, experiences, and specialized expertise are better positioned to recombine existing knowledge, solve complex technical problems, explore alternative solution paths, and generate novel ideas (Ancona and Caldwell 1992, Taylor and Greve 2006, Choudhury and Haas 2018, Ferrucci and Lissoni 2019).

To explore it, we measure the collective skill capacity of a development team as the number of distinct skills represented among its members. As shown in column 2 of Table 8, remote work adoption is associated with a significant increase in the number of distinct skills within app development teams. This finding is consistent with the argument that remote work enhances the collective skill capacity of development teams, which may in turn strengthen development capability and digital product innovation performance.

[insert Table 8 here]

Because identifying the mechanisms underlying the estimated effects is more challenging than estimating the effects themselves (Forderer and Burtch 2025), we provide exploratory analyses examining the potential roles of team size and collective skill capacity in shaping app innovation performance in Online Appendix D. These analyses are at best suggestive of the mechanisms at work. Using the LA-PSM matched sample, we find that both team size and collective skill capacity are positively and significantly associated with the number of major releases and the number of new features. Furthermore, the estimated effect of remote work adoption is attenuated when these variables are included in the model. These results provide suggestive evidence that workforce expansion and enhanced collective skill capacity may represent important pathways through which remote work influences app innovation performance.

#### 6.4.3 Alternative Explanations

Remote work has been shown to improve individual productivity by reducing commuting time (Choudhury et al. 2021). One potential alternative explanation for our findings is therefore that the observed increase in app innovation performance is driven primarily by commuting-time reductions. To examine whether the reduced commuting time translates into greater app innovation performance, we use traffic volume by the U.S. Census Tract Data to measure average traffic congestion in the cities where individuals performing development tasks are located (Finlay et al. 2022). Due to data availability, we restrict the analysis to U.S.-based development teams. We classify apps into low-traffic and high-traffic groups based on whether the average congestion level in their city prior to remote work adoption falls below or above the sample median, where higher traffic levels correspond to longer commuting times. We then re-estimate our main model separately for the two groups. Table 9 reports the results. We find that remote work adoption is associated with improved app innovation performance in both groups, and the estimated effects do not differ significantly between high-traffic and low-traffic locations. These findings suggest that reductions in commuting time are unlikely to be the primary driver of the app innovation improvements associated with remote work.

[insert Table 9 here]

Another potential explanation is that the impact of remote work may depend on the maturity of the app being developed. More mature apps may benefit differently from remote work than younger apps because they typically differ in development priorities, technical complexity, and innovation opportunities. To examine this possibility, we classify apps into young and old groups based on app age using a median split and re-estimate our main model separately for each group. Table 10 reports the results. We find that remote work adoption is associated with higher app innovation performance for both young and old apps, and the estimated effects do not differ significantly across the two groups. These findings suggest that differences in app maturity are unlikely to explain the positive relationship between remote work and app innovation performance.

[insert Table 10 here]

## 7. Discussion and Conclusion

As remote work becomes increasingly prevalent with advances in digital technologies, it is important to understand its implications for team related activities such as continuous digital product innovation. Using mobile app development as our empirical context, we examine how remote work adoption influences app innovation performance. We find that remote work increased both major releases and new feature introductions, suggesting that the productivity gains associated with remote work outweigh the coordination challenges that arise when development teams transition to remote work arrangements. Importantly, these gains did not come at the expense of innovation originality. Moreover, the observed improvements in app innovation performance were accompanied by increased user demand, indicating that remote work can generate meaningful market benefits in addition to enhancing product innovation outcomes. To explore potential explanations, we find that development teams with prior modular collaboration experience realized larger app innovation gains following remote work adoption, suggesting that experience coordinating work in modular environments helps development teams manage distributed collaboration more effectively and further mitigates coordination challenges. We also find that remote work led to increases in both team size and collective skill capacity, providing evidence that access to a broader talent pool expands development resources and capabilities. These changes are associated with improved app innovation performance and represent plausible pathways through which remote work affects app innovation outcomes. Finally, we find no evidence that the effects are driven primarily by reductions in commuting time or differences in app maturity, helping to rule out these alternative explanations for the observed benefits of remote work.

Our study has several theoretical implications. First, it contributes to the emerging literature on remote work and innovation by suggesting that the effects of remote work depend critically on the nature of the innovation process. Prior studies have highlighted both the potential productivity benefits of remote work (Bloom et al. 2015, Choudhury et al. 2021, Angelici and Profeta 2024, Bloom et al. 2024) and the coordination challenges arising from reduced face-to-face interaction and communication (Brucks and Levav 2022, Yang et al. 2022, Lin et al. 2023). Our results indicate that, in digital product development, where innovation is continuous, modular, and supported by digital collaboration tools, the benefits of remote work can outweigh its coordination costs. This finding suggests that innovation consequences of remote work depend on the characteristics of the innovation process, particularly the extent to which innovation activities can be modularized and coordinated through digital technologies.

Second, our findings highlight the role of organizational coordination capabilities in the effectiveness of remote work. Prior research has long emphasized that coordination capabilities, shared routines, and modular organizational structures facilitate knowledge integration across distributed contributors (Majchrzak et al. 2005, Enkel et al. 2009, Kleis et al. 2012). Research on modularity similarly argues that modular architectures reduce coordination requirements and enable distributed innovation (Baldwin and Clark 2000, MacCormack et al. 2006). We extend this literature by showing that prior modular collaboration experience enables development teams to realize greater digital product innovation benefits from remote work. Rather than viewing remote work as uniformly beneficial or detrimental, our findings suggest that its performance consequences depend on organizations' existing coordination capabilities and their ability to manage distributed collaboration effectively.

Third, our findings provide new evidence that remote work influences innovation not only by changing how work is coordinated but also by reshaping firms' access to human capital. Prior studies have shown that remote work expands geographic hiring and improves firms' access to talent (Ham et al. 2022, Choudhury 2025, Hsu and Tambe 2025). We extend this literature by demonstrating that remote work policy leads to bigger team size and increased skill capacity and these can improve downstream digital product innovation. More broadly, our findings suggest that remote work should be viewed as an organizational design choice that jointly affects coordination processes and the allocation of human capital. This perspective connects the remote work literature with broader theories of organizational design, which emphasize that organizational performance depends on aligning work arrangements, coordination mechanisms, and human capital with the underlying production process.

Our findings also have several managerial implications. First, organizations should view remote work as one component of a broader organizational design rather than as an isolated human resource policy. The effectiveness of remote work depends on whether complementary organizational practices are in place to support distributed collaboration. Managers should therefore evaluate remote work together with decisions about work processes, coordination mechanisms, and team structure rather than treating workplace flexibility as an independent policy choice.

Second, organizations should invest in developing coordination capabilities before expanding remote work arrangements. Our findings suggest that the benefits of remote work depend on teams' ability to coordinate modular and distributed work. Managers can strengthen these capabilities by encouraging modular development practices, establishing standardized workflows, and creating collaboration routines that support geographically dispersed teams. Building these organizational capabilities can enable teams to better adapt to distributed work and sustain digital product innovation performance over time.

Third, organizations should leverage remote work as part of a broader talent strategy rather than simply as a flexible work arrangement. Remote work enables organizations to recruit beyond local labor markets, expanding both the size and skill capacity of development teams. Managers should therefore evaluate remote work not only by its impact on employees but also by its potential to broaden access to specialized expertise. More broadly, remote work provides organizations with an opportunity to rethink how talent is sourced, organized, and coordinated to support continuous digital product innovation.

Our study has several limitations. First, we do not directly observe internal decisions regarding remote work adoption. Instead, we infer adoption timing from remote work policies disclosed in job postings. Although our instrumental variable strategy mitigates concerns about measurement error, future research with access to internal policy records could provide more precise evidence. Second, we do not observe qualitative factors such as work-life-balance, job satisfaction, or motivation, which may influence the relationship between remote work and digital product innovation outcomes. Future research incorporating such measures could further illuminate these channels. Third, because our sample focuses on the top 500 apps in each category, our findings may not apply to unsuccessful or discontinued digital products. Future research could examine whether remote work influences product innovation output differently for failed or lower-performing digital products. Future studies should also validate the findings for other type of digital products.

**Tables and Figures**

Table 1: Definitions and Summary Statistics, App-Month-Level

| Variables | Description | Obs | Mean | Std. | Min | Max |
|---|---|---|---|---|---|---|
| Major_releases | Number of major releases per app | 56,213 | 9.06 | 13.79 | 0 | 53 |
| New_features | Number of new features per app | 56,213 | 40.50 | 50.63 | 1 | 188 |
| Original_new_features | Number of original new features per app | 56,213 | 34.59 | 47.04 | 0 | 174 |
| Copycat_new_features | Number of copycat new features per app | 56,213 | 5.74 | 4.95 | 0 | 18 |
| Copycat_new_features_ratio | Share of copycat new features per app | 56,166 | 0.38 | 0.44 | 0 | 1 |
| Download | Number of download volume per app | 56,213 | 106,768 | 232,002 | 65.08 | 932,646 |
| Same_skill_labor | The number of employees associated with pre-existing fixed role–skill combination in the focal app development team | 56,213 | 151.73 | 195.05 | 3 | 742 |
| Skill_capacity | The total number of distinct skills in the focal app development team | 56,213 | 1465.35 | 1386.39 | 70 | 5041 |

Table 2: The Impact of Remote Work on App Innovation Performance

| | (1)<br>Log (major_releases) | (2)<br>Log (new_features) |
|---|---|---|
| ATT of WFH | 0.100*** | 0.111*** |
| | (0.021) | (0.020) |
| App FE | Yes | Yes |
| Month FE | Yes | Yes |
| # of apps | 1,782 | 1,782 |
| Observations | 35,108 | 35,108 |

Robust standard errors clustered at app level in parentheses. *** p<0.01, ** p<0.05, * p<0.1.

Table 3: The Impact of Remote Work on App Major Releases (Robustness Checks)

| | Treatment Threshold | Window Size 18 months | Window Size 6 months | LA-PSM and TWFE | Post-pandemic Analysis |
|---|---|---|---|---|---|
| | (1)<br>Log (major_release) | (2)<br>Log (major_release) | (3)<br>Log (major_release) | (4)<br>Log (major_release) | (5)<br>Log (major_release) |
| ATT of WFH | 0.049*** | 0.059** | 0.045** | 0.025*** | 0.083** |
| | (0.016) | (0.028) | (0.020) | (0.007) | (0.039) |
| App FE | Yes | Yes | Yes | Yes | Yes |
| Month FE | Yes | Yes | Yes | Yes | Yes |
| # of apps | 1,640 | 1,782 | 1,782 | 1,638 | 718 |
| Observations | 33,208 | 45,730 | 20,536 | 51,419 | 12,706 |

Robust standard errors clustered at app level in parentheses. *** p<0.01, ** p<0.05, * p<0.1.

Table 4: The Impact of Remote Work on App New Features (Robustness Checks)

| | Treatment Threshold | Window Size 18 months | Window Size 6 months | LA-PSM and TWFE | Post-pandemic Analysis |
|---|---|---|---|---|---|
| | (1)<br>Log (new_features) | (2)<br>Log (new_features) | (3)<br>Log (new_features) | (4)<br>Log (new_features) | (5)<br>Log (new_features) |
| ATT of WFH | 0.055*** | 0.084*** | 0.049** | 0.032*** | 0.103*** |
| | (0.018) | (0.026) | (0.021) | (0.006) | (0.037) |
| App FE | Yes | Yes | Yes | Yes | Yes |
| Month FE | Yes | Yes | Yes | Yes | Yes |
| # of apps | 1,640 | 1,782 | 1,782 | 1,638 | 718 |
| Observations | 33,208 | 45,730 | 20,536 | 51,419 | 12,706 |

Robust standard errors clustered at app level in parentheses. *** p<0.01, ** p<0.05, * p<0.1.

Table 5: The Impact of Remote Work on App Innovation Performance (IV Analysis)

| First-stage: | (1) A's WFH | Second-stage: | (1) Log (major_releases) | (2) Log (new_features) |
|---|---|---|---|---|
| C's Internet Connectivity | 2.148*** | ATT of WFH | 0.287*** | 0.130** |
| | (0.249) | | (0.088) | (0.055) |
| F-statistics | 74.41 | | | |
| B's labor controls | Yes | B's labor controls | Yes | Yes |
| A's Location FE | Yes | A's Location FE | Yes | Yes |
| App FE | Yes | App FE | Yes | Yes |
| Month FE | Yes | Month FE | Yes | Yes |
| # of apps | 633 | # of apps | 633 | 633 |
| Observations | 20,941 | Observations | 20,941 | 20,941 |

Robust standard errors clustered at app level in parentheses. *** p<0.01, ** p<0.05, * p<0.1.

Table 6: The Impact of Remote Work on App Innovation Originality and App Demand

| | (1) Log (original_new_features) | (2) Log (copycat_new_features) | (3) Copycat_new_features_ratio | (4) Log (download) |
|---|---|---|---|---|
| ATT of WFH | 0.109*** | 0.079*** | 0.013 | 0.307*** |
| | (0.027) | (0.011) | (0.010) | (0.043) |
| App FE | Yes | Yes | Yes | Yes |
| Month FE | Yes | Yes | Yes | Yes |
| # of apps | 1,782 | 1,782 | 1,782 | 1,782 |
| Observations | 35,108 | 35,108 | 35,108 | 35,108 |

Robust standard errors clustered at app level in parentheses. *** p<0.01, ** p<0.05, * p<0.1.

Table 7: The Impact of Remote Work on App Innovation Performance
(OSS vs. non-OSS App Development Teams)

| | OSS | Non-OSS | Diff | OSS | Non-OSS | Diff |
|---|---|---|---|---|---|---|
| | (1) Log (major_release) | (2) Log (major_release) | (3) Log (major_release) | (4) Log (new_features) | (5) Log (new_features) | (6) Log (new_features) |
| ATT of WFH | 0.229*** | 0.078*** | 0.151** | 0.242*** | 0.090*** | 0.152** |
| | (0.066) | (0.022) | (0.073) | (0.074) | (0.024) | (0.075) |
| App FE | Yes | Yes | | Yes | Yes | |
| Month FE | Yes | Yes | | Yes | Yes | |
| # of apps | 369 | 1,413 | | 369 | 1,413 | |
| Observations | 7,034 | 28,074 | | 7,034 | 28,074 | |

Robust standard errors clustered at app level in parentheses. *** p<0.01, ** p<0.05, * p<0.1.

Table 8: The Impact of Remote Work on App Development Team Size and Skills

| | (1) Log (same_skill_labor) | (2) Log (skill_capacity) |
|---|---|---|
| ATT of WFH | 0.261*** | 0.228*** |
| | (0.024) | (0.021) |
| App FE | Yes | Yes |
| Month FE | Yes | Yes |
| # of apps | 1,782 | 1,782 |
| Observations | 35,108 | 35,108 |

Robust standard errors clustered at app level in parentheses. *** p<0.01, ** p<0.05, * p<0.1.

Table 9: The Impact of Remote Work on App Innovation Performance
(Low-Traffic vs. High-Traffic App Development Teams)

| | High Traffic | Low Traffic | Diff | High Traffic | Low Traffic | Diff |
|---|---|---|---|---|---|---|
| | (1) | (2) | (3) | (4) | (5) | (6) |
| | Log (major_release) | Log (major_release) | Log (major_release) | Log (new_features) | Log (new_features) | Log (new_features) |
| ATT of WFH | 0.129*** | 0.134*** | -0.005 | 0.143*** | 0.139*** | 0.004 |
| | (0.046) | (0.037) | (0.055) | (0.045) | (0.037) | (0.062) |
| App FE | Yes | Yes | | Yes | Yes | |
| Month FE | Yes | Yes | | Yes | Yes | |
| # of apps | 608 | 654 | | 608 | 654 | |
| Observations | 11,915 | 12,569 | | 11,915 | 12,569 | |

Note: The estimation sample is smaller than that in the main analysis because the traffic data is available only for U.S.-based development teams. Accordingly, the analysis is restricted to U.S.-based development teams and their corresponding apps. Robust standard errors clustered at app level in parentheses. *** p<0.01, ** p<0.05, * p<0.1.

Table 10: The Impact of Remote Work on App Innovation Performance
(Young vs. Old Apps)

| | Old App | Young App | Diff | Old App | Young App | Diff |
|---|---|---|---|---|---|---|
| | (1) | (2) | (3) | (4) | (5) | (6) |
| | Log (major_release) | Log (major_release) | Log (major_release) | Log (new_features) | Log (new_features) | Log (new_features) |
| ATT of WFH | 0.045** | 0.053*** | -0.008 | 0.045*** | 0.074*** | -0.029 |
| | (0.021) | (0.014) | (0.027) | (0.017) | (0.017) | (0.024) |
| App FE | Yes | Yes | | Yes | Yes | |
| Month FE | Yes | Yes | | Yes | Yes | |
| # of apps | 887 | 895 | | 887 | 895 | |
| Observations | 17,397 | 17,711 | | 17,397 | 17,711 | |

Robust standard errors clustered at app level in parentheses. *** p<0.01, ** p<0.05, * p<0.1.

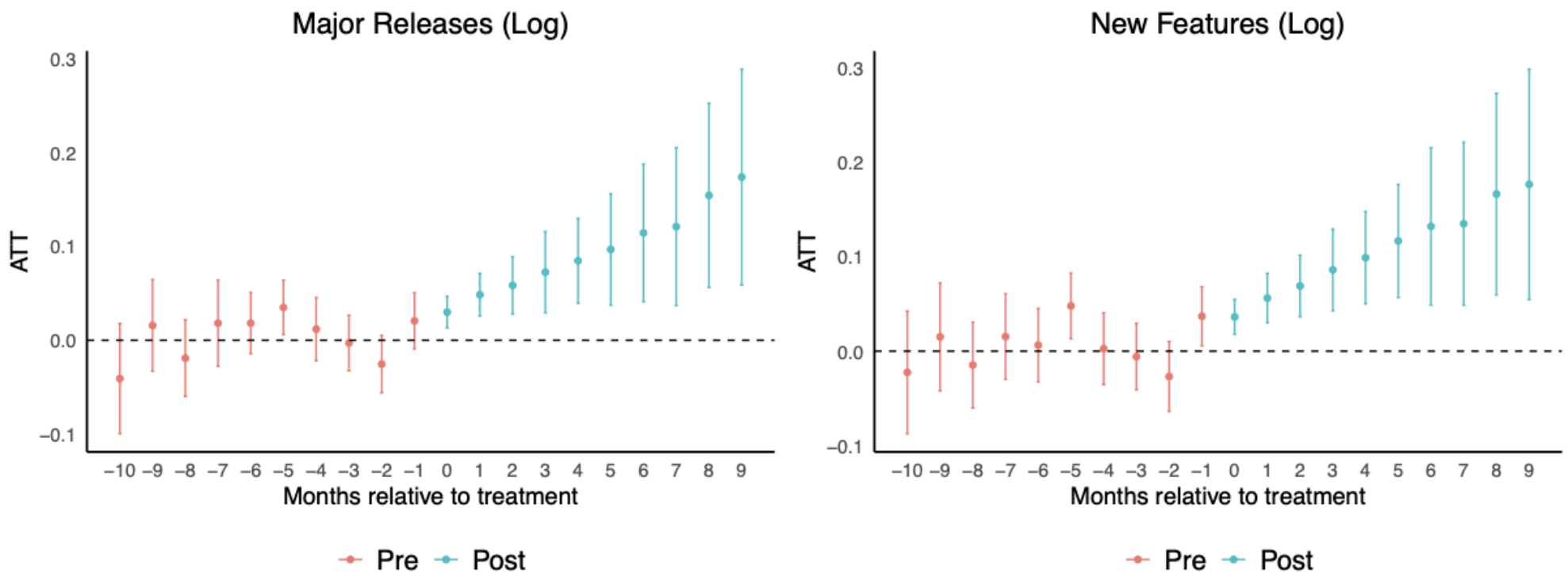

Figure 1: Parallel Trend Tests of App Innovation Performance

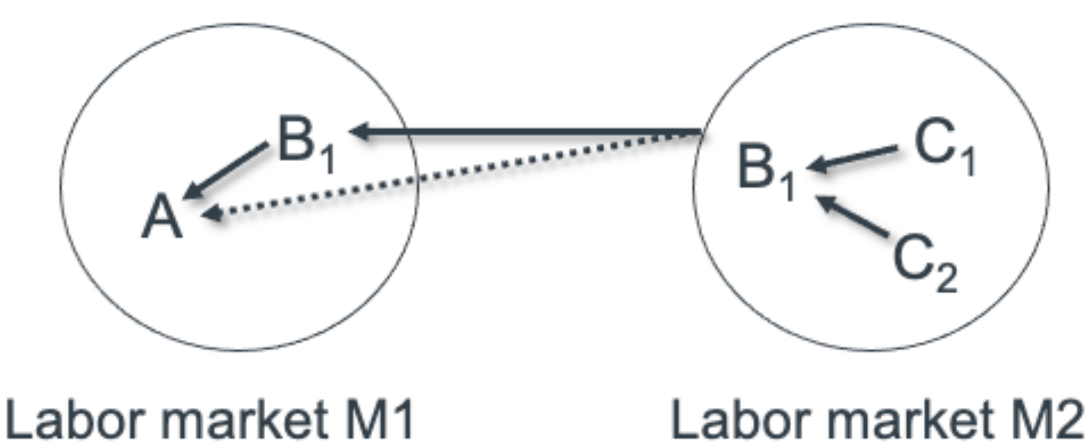

Figure 2: Indirect Effect of Labor Competitors' Competitors' Internet Connectivity on the Remote Work Decisions of Focal App Development Team

## Online Appendix

### Appendix A. Detection of Remote Work Policy

To identify the adoption of remote work and its timing, we follow established approaches in the economics literature that infer work arrangement policies (Bloom et al. 2021). Specifically, we search job titles and job descriptions for a comprehensive set of work-from-home (WFH)-related keywords that capture a wide range of terminology commonly used to describe remote and flexible work arrangements.

Based on prior literature (Bloom et al. 2021), we compile the following set of WFH-related keywords: telecommuting, tele-commuting, telework, teleworking, working from home, work from home, work-from-home, mobile work, mobile working, remote work, remote working, work remotely, working remotely, remote workplace, flexible workplace, flexible work, work-at-home, work at home, telecommuter, tele-worker, telecommuting specialist, home-sourced worker, home-sourced employee, home office, home-based office, home-based work, nomadic worker, nomadic employee, work from anywhere, working from anywhere, work-from-anywhere, digital workplace, virtual work, virtual office, virtual employee, distance work, working from a remote location, work from a remote location, video conference, video conferencing, video chat, video call, teleconference, and teleconferencing. These keywords are used to identify job postings that potentially reference remote or flexible work arrangements.

The presence of WFH-related keywords alone is insufficient to infer remote work adoption, as such language may also be used to describe restrictive policies (e.g., statements explicitly prohibiting remote work). To address this issue, we implement a rule-based text-classification procedure that evaluates the local linguistic context surrounding each keyword occurrence. This procedure distinguishes between permissive references that indicate remote work is allowed and restrictive references that indicate remote work is prohibited or unavailable. We use permissive references to identify the presence of remote work.

**Appendix B. Release and Feature Classification Details**

**B.1 Release Classification**

To distinguish substantive innovation-oriented updates from routine maintenance updates, we classify each app version release as either a *major release* or a *minor release* using a large language model (LLM) based classification approach (Brynjolfsson et al. 2025). Specifically, we apply OpenAI's GPT-5.5 model to the release descriptions of 98,512 app versions in our sample (Huang et al. 2026). This approach allows us to classify software update logs at scale while preserving sensitivity to the nuanced language commonly used in release notes. Consistent with established practices in the literature (Foerderer et al. 2018), we define a *major release* as an update that introduces substantial changes to the product, such as new functionality, interface redesigns, major integrations, or key enhancements to user workflows. In contrast, a *minor release* refers to an update primarily involving bug fixes, stability improvements, performance tuning, or other incremental changes that do not materially alter the app's core functionality or user experience.

For each version description, we submit the release-note text to GPT-5.5 using the following prompt:

*You are an expert evaluating iOS application version release descriptions. Your task is to classify each release as either a* ***major release*** *or* ***minor release*** *based on the following criteria:*

*Definitions:*
*Major release:*
*A release that introduces substantial changes to the application's functionality or user experience.*
*These changes typically include one or more of the following:*
*- Introduction of entirely new functionality or capabilities*
*- Major interface redesigns or workflow changes*
*- New system integrations or platform-level capabilities*
*- Structural or architectural changes that meaningfully expand the app's capabilities*

*Minor release:*
*A release focused primarily on maintenance or incremental improvements that do not substantially change the app's core functionality.*
*These typically include:*
*- Bug fixes*

*- Stability improvements*
*- Small UI tweaks*
*- Compatibility updates*

*Output format:*
*Carefully analyze the description. Use step-by-step reasoning to explain your thinking. Return only one of the following labels on the final line:*
*Final classification: Major*
*or*
*Final classification: Minor*

*Release description:*
*""" description """*

To assess the validity of the GPT-based classification, we randomly selected 100 release descriptions from our sample and asked independent human evaluators to classify each release as either major or minor using the same conceptual criteria. The evaluators performed the classification without access to the GPT-generated labels. The human and GPT classifications were consistent in 94 of the 100 cases, yielding an agreement rate of 94%.

**B.2 Feature Classification**

To identify whether a feature introduced in a release represents a *new* functionality or an *existing* one, we analyze the semantic content of release-note text using large language model (LLM)–based embeddings. Specifically, we use OpenAI's text-embedding-3-large model to generate textual embeddings for each feature description extracted from the release notes.

Textual embeddings convert a piece of text into a high-dimensional vector representation that captures its semantic meaning (Brynjolfsson et al. 2025). In this representation, two pieces of text that describe similar functionality will occupy nearby locations in the semantic space, whereas text describing different functionality will be located further apart. The exact coordinates depend on the embedding model used. The text-embedding-3-large model is designed to capture semantic relationships between text segments and allow conceptually similar descriptions to be grouped together even when they use different wording.

For each feature description, we generate a vector embedding using the OpenAI's text-embedding-3-large model. We then compare the semantic similarity between features introduced in the current version and features appearing in prior versions of the same app. Similarity between two feature descriptions is measured using cosine similarity between their corresponding embedding vectors. Cosine similarity is commonly used to compare semantic embeddings and ranges from 0, indicating that two pieces of text are semantically unrelated, to 1, indicating that they convey identical meaning (Koroteev 2021).

Our goal is to identify whether a feature introduces terminology that has not appeared in any previous release of the same app. Because release notes often describe the same functionality using different wording across versions, simple keyword matching may incorrectly classify existing features as new when the wording changes. To address this issue, we use semantic similarity rather than exact keyword matching to determine whether a feature reuses previously observed terminology. Specifically, we compute the cosine similarity between the embedding of the current feature description and the embeddings of all feature descriptions observed in prior releases of the same app. If the maximum cosine similarity exceeds a predefined similarity threshold, we treat the current description as semantically reusing previously observed terminology and classify the feature as existing. Conversely, if the similarity between the current feature embedding and all previously observed feature embeddings falls below the threshold, we classify the feature as new, indicating the introduction of a functionality whose terminology has not previously appeared in the app's release history. We set the similarity threshold to 0.5, such that a feature is classified as new only when its description exhibits relatively low semantic similarity to all previously observed feature descriptions. This conservative threshold reduces the likelihood that semantically related functionalities described using different wording are incorrectly identified as novel features. The threshold was selected based on calibration against a manually labeled validation sample to balance false matches and missed matches in identifying feature novelty.

To validate the classification procedure, we randomly sampled 100 feature descriptions and manually evaluated whether each feature represented a new or existing functionality based on the full release history of the app. The human classifications matched the LLM-based classifications in 92% of the cases, suggesting that the embedding-based approach provides a reliable method for identifying feature novelty.

## Appendix C. Additional Robustness Checks

Table C.1: Parallel Trends Test (LA-PSM Sample)

| | (1)<br>Log (major_releases) | (2)<br>Log (new_features) |
|---|---|---|
| $Treatment_i \times Pre4m$ | 0.007 | 0.010 |
| | (0.007) | (0.007) |
| $Treatment_i \times Pre3m$ | 0.010 | 0.009 |
| | (0.007) | (0.008) |
| $Treatment_i \times Pre2m$ | 0.010 | 0.007 |
| | (0.008) | (0.009) |
| $Treatment_i \times Pre1m$ | 0.011 | 0.012 |
| | (0.009) | (0.009) |
| $Treatment_i \times Post0m$ | 0.026*** | 0.029*** |
| | (0.009) | (0.010) |
| $Treatment_i \times Post1m$ | 0.029*** | 0.034*** |
| | (0.010) | (0.010) |
| $Treatment_i \times Post2m$ | 0.028*** | 0.035*** |
| | (0.010) | (0.010) |
| $Treatment_i \times Post3m$ | 0.031*** | 0.037*** |
| | (0.010) | (0.011) |
| $Treatment_i \times Post4m$ | 0.033*** | 0.039*** |
| | (0.011) | (0.011) |
| $Treatment_i \times Post5m +$ | 0.039*** | 0.049*** |
| | (0.012) | (0.012) |

Robust standard errors clustered at app level in parentheses. *** p<0.01, ** p<0.05, * p<0.1.

Table C.2: The Impact of Remote Work on App Innovation Performance (Alternate Measure)

| | (1)<br>Log (major_versions) |
|---|---|
| ATT of WFH | 0.021** |
| | (0.009) |
| App FE | Yes |
| Month FE | Yes |
| # of apps | 1,441 |
| Observations | 28,174 |

Robust standard errors clustered at app level in parentheses. *** p<0.01, ** p<0.05, * p<0.1.

Table C.3: The Impact of Remote Work on App Innovation Performance (WFH Policy Fluctuation)

| | (1)<br>Log (major_releases) | (2)<br>Log (new_features) |
|---|---|---|
| ATT of WFH | 0.107*** | 0.113*** |
| | (0.024) | (0.022) |
| App FE | Yes | Yes |
| Month FE | Yes | Yes |
| # of apps | 1,782 | 1,782 |
| Observations | 33,004 | 33,004 |

Robust standard errors clustered at app level in parentheses. *** p<0.01, ** p<0.05, * p<0.1.

Table C.4: The Impact of Remote Work on App Innovation Performance
(Lagged Effects)

| | Monthly Level | | Quarterly Level | |
|---|---|---|---|---|
| | (1) | (2) | (3) | (4) |
| | Log (major_releases) | Log (new_features) | Log (major_releases) | Log (new_features) |
| ATT of WFH | 0.086*** | 0.097*** | 0.051** | 0.062*** |
| | (0.022) | (0.021) | (0.022) | (0.022) |
| App FE | Yes | Yes | Yes | Yes |
| Month FE | Yes | Yes | Yes | Yes |
| # of apps | 1,782 | 1,782 | 1,471 | 1,471 |
| Observations | 34,807 | 34,807 | 10,888 | 10,888 |

Robust standard errors clustered at app level in parentheses. *** p<0.01, ** p<0.05, * p<0.1.

Table C.5: The Impact of Remote Work on App Innovation Performance
(Firm-level Performance)

| | (1) | (2) |
|---|---|---|
| | Log (major_releases) | Log (new_features) |
| ATT of WFH | 0.165*** | 0.183*** |
| | (0.031) | (0.030) |
| Firm FE | Yes | Yes |
| Month FE | Yes | Yes |
| # of firms | 937 | 937 |
| Observations | 18,280 | 18,280 |

Robust standard errors clustered at firm level in parentheses. *** p<0.01, ** p<0.05, * p<0.1.

**Appendix D. Team Size, Skill Capacity and App Innovation Performance**

Table D.1: Team Size and App Innovation Performance

| | (1) Log (major_releases) | (2) Log (new_features) | (3) Log (major_releases) | (4) Log (new_features) | (5) Log (major_releases) | (6) Log (new_features) |
|---|---|---|---|---|---|---|
| ATT of WFH | 0.025*** | 0.032*** | | | 0.023*** | 0.030*** |
| | (0.007) | (0.006) | | | (0.007) | (0.006) |
| Log (same_skill_labor) | | | 0.058** | 0.049** | 0.053** | 0.043* |
| | | | (0.024) | (0.023) | (0.024) | (0.023) |
| App FE | Yes | Yes | Yes | Yes | Yes | Yes |
| Month FE | Yes | Yes | Yes | Yes | Yes | Yes |
| # of apps | 1,638 | 1,638 | 1,638 | 1,638 | 1,638 | 1,638 |
| Observations | 51,419 | 51,419 | 51,419 | 51,419 | 51,419 | 51,419 |

Robust standard errors clustered at app level in parentheses. *** p<0.01, ** p<0.05, * p<0.1.

Table D.2: Team Collective Skills and App Innovation Performance

| | (1) Log (major_releases) | (2) Log (new_features) | (3) Log (major_releases) | (4) Log (new_features) | (5) Log (major_releases) | (6) Log (new_features) |
|---|---|---|---|---|---|---|
| ATT of WFH | 0.025*** | 0.032*** | | | 0.022*** | 0.029*** |
| | (0.007) | (0.006) | | | (0.007) | (0.006) |
| Log (skill_capacity) | | | 0.050** | 0.053*** | 0.044** | 0.046** |
| | | | (0.022) | (0.020) | (0.022) | (0.020) |
| App FE | Yes | Yes | Yes | Yes | Yes | Yes |
| Month FE | Yes | Yes | Yes | Yes | Yes | Yes |
| # of apps | 1,638 | 1,638 | 1,638 | 1,638 | 1,638 | 1,638 |
| Observations | 51,419 | 51,419 | 51,419 | 51,419 | 51,419 | 51,419 |

Robust standard errors clustered at app level in parentheses. *** p<0.01, ** p<0.05, * p<0.1.